\documentclass[twocolumn, prl, amssymb, superscriptaddress, aps, preprintnumbers,amsmath,floatfix]{revtex4} 
\usepackage{multirow}
\usepackage{isotope}
\usepackage{graphicx}
\usepackage{color}

\newcommand{\rket}[1]{\left\| #1 \right\rangle}
\newcommand{\rbra}[1]{\left\langle #1 \right\|}

\newcommand{\cf}[1]{\hat{\mathcal{#1}}}

\newcommand{\ttau}{\lambda}

\begin{document}

\title{Electronic bridge excitation in highly charged $\isotope[229]{Th}$ ions}

\author{Pavlo V. Bilous}
\email{Pavlo.Bilous@mpi-hd.mpg.de}
\affiliation{Max-Planck-Institut f\"ur Kernphysik, Saupfercheckweg 1, D-69117 Heidelberg, Germany}

\author{Hendrik Bekker}
\affiliation{Max-Planck-Institut f\"ur Kernphysik, Saupfercheckweg 1, D-69117 Heidelberg, Germany}
\affiliation{Department of Physics, Columbia University, 538 West 120th Street, New York, NY, 10027-5255, USA}

\author{Julian Berengut}
\affiliation{University of New South Wales, Sydney NSW 2052, Australia}
\affiliation{Max-Planck-Institut f\"ur Kernphysik, Saupfercheckweg 1, D-69117 Heidelberg, Germany}

\author{Benedict Seiferle}
\affiliation{Ludwig-Maximilians-Universit\"at M\"unchen, Am Coulombwall 1,                              
D-85748 Garching, Germany}

\author{Lars von der Wense}
\affiliation{Ludwig-Maximilians-Universit\"at M\"unchen, Am Coulombwall 1,                              
D-85748 Garching, Germany}

\author{Peter G. Thirolf}
\affiliation{Ludwig-Maximilians-Universit\"at M\"unchen, Am Coulombwall 1,                              
D-85748 Garching, Germany}

\author{Thomas Pfeifer}
\affiliation{Max-Planck-Institut f\"ur Kernphysik, Saupfercheckweg 1, D-69117 Heidelberg, Germany}

\author{Jos\'{e} R. Crespo L\'{o}pez-Urrutia}
\affiliation{Max-Planck-Institut f\"ur Kernphysik, Saupfercheckweg 1, D-69117 Heidelberg, Germany}

\author{Adriana P\'alffy}
\email{Palffy@mpi-hd.mpg.de}
\affiliation{Max-Planck-Institut f\"ur Kernphysik, Saupfercheckweg 1, D-69117 Heidelberg, Germany}


\date{\today}

\begin{abstract}
The excitation of the 8 eV $^{229m}$Th isomer through the electronic bridge mechanism in highly charged ions is investigated theoretically. By exploiting the rich level scheme of open $4f$ orbitals and the robustness of highly charged ions against photoionization, a pulsed high-intensity optical laser can be used to efficiently drive the nuclear transition by coupling it to the electronic shell. We show how to implement a promising electronic bridge scheme in an electron beam ion trap starting from a metastable electronic state. This setup would avoid the need for a tunable vacuum ultraviolet laser. Based on our theoretical predictions, determining the isomer energy with an uncertainty of $10^{-5}$~eV could be achieved in one day of measurement time using realistic laser parameters.

\end{abstract}

\maketitle

Throughout the entire nuclear chart, the nuclear excitation with the lowest known energy occurs in the actinide region in $\isotope[229]{Th}$. Its first excited state lies at only 8~eV and is a nuclear isomer, i.~e., a long-lived nuclear state. From the nuclear structure point of view, this peculiar state appears due to a unique very fine interplay between collective and single-particle degrees of freedom \cite{Minkov_Palffy_PRL_2017}. More importantly, there is a strong interdisciplinary interest for $\isotope[229m]{Th}$ arising from its possible applications. 
Future vacuum-ultraviolet (VUV) laser access to the $\isotope[229]{Th}$ isomer promises a new ``nuclear clock'' frequency standard \cite{Peik_Clock_2003,Campbell_Clock_2012,Peik_Clock_2015},
the development of nuclear lasers in the optical range \cite{Tkalya_NuclLaser_2011} as well as detection and precise determination of the temporal variation of fundamental constants
\cite{Flambaum06,Berengut_ConstVar_PRL_2009,Rellegert2010}. Practical implementations will require both a precise knowledge of the isomer energy $E_m$ as well as means to drive the isomeric transition.

\begin{figure}[ht!]
\centering
\includegraphics[width=0.43\textwidth]{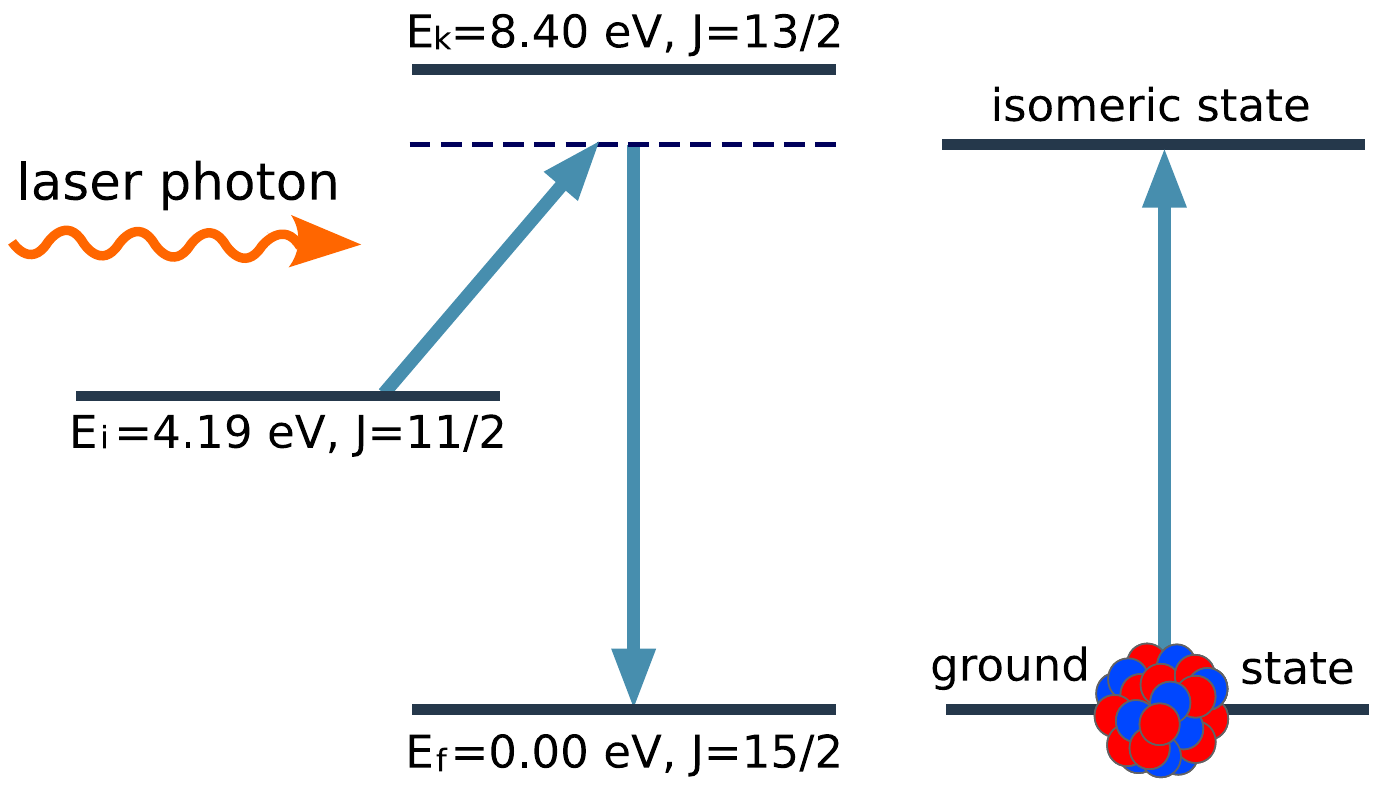}
 \caption{Simplified scheme of the EB excitation mechanism. The electronic shell (left) initially in an excited state  $E_i$ is promoted by a laser photon to a virtual state (dashed line), which decays to a lower-lying real state $E_f$ by transferring its energy to the nucleus. The latter undergoes the transition from the ground to the isomeric state (right). 
The electronic states are labeled by their energy and angular momentum $J$.}
\label{EB_scheme}
\end{figure}

In a recent experiment, a direct measurement of the internal conversion (IC) electrons emitted in the isomer decay allowed for the most accurate energy determination thus far, yielding $E_m=8.28 \pm 0.17$~eV \cite{Seiferle_Nature_2019}. The controlled excitation of the isomer has been so far achieved only via  x-ray pumping of higher levels with synchrotron radiation  \cite{Japanese_Nature_2019}. Other proposed approaches include direct photoexcitation of Th in a solid-state environment with a tunable VUV laser \cite{Wense_PRL_2017} or  excitation through the coupling of the isomer to the electronic shell in a so called electronic bridge (EB) process  \cite{TkalyaBridge1992, PorsevFlambaum_Brige_PRL_2010}. The EB mechanism employs a tunable optical or UV laser to drive the electronic shell. Via a virtual electronic state, the atomic excitation energy is transferred to the nucleus and the isomer populated as illustrated in Fig.~\ref{EB_scheme}.
 So far, EB schemes for excitation and decay have been investigated for Th$^{+}$ \cite{PorsevFlambaum_Brige1+_PRA_2010}, Th$^{2+}$ \cite{Peik_Clock_2015}, Th$^{3+}$ \cite{PorsevFlambaum_Brige3+_PRA_2010,Bilous_NJP_2018} and recently for $\isotope[235]{U}^{7+}$ \cite{Berengut_PRL_2018}. Highly charged ions (HCI) of Th such as Th$^{89+}$ have only been investigated in the context of precise hyperfine splitting calculations \cite{Karpeshin_1s_PRC_1998,Tkalya_1s_PRC_2016}. In Th$^{+}$--Th$^{3+}$ ions, the EB approach depends on the availability of magnetic dipole ($M1$) or electric quadrupole ($E2$) electronic transitions that match the multipolarity of the nuclear transition \cite{Bilous_PRC_2018}. This can be a strong limitation due to the incomplete knowledge on the $\isotope{Th}^+$ electronic levels at excitation energies higher than 8~eV \cite{Meier_PRA_2019, HerreraSancho_PRA_2012}. Moreover, the applied laser power (which enters the EB rate as a multiplicative factor) is limited due to possible multiphoton ionization.

In this Letter, we investigate the EB mechanism in Th HCI, which are very robust against photoionization and can handle high laser intensities. We focus in particular on HCI with an open $4f$-shell, which possess a rich choice of $M1$ and $E2$ electronic transitions within the fine structure of the ground state configuration. For the particular case of the $\isotope{Th}^{35+}$ ion (electronic configuration $[\isotope{Kr}]\,4d^{10} 4f^9 $) we show that excitation of the $\isotope[229m]{Th}$ isomer can be achieved by means of a standard tunable UV laser, provided its energy lies within the predicted error bars $E_m=8.28 \pm 0.17$~eV \cite{Seiferle_Nature_2019}. The EB excitation would at the same time  allow for the determination of the energy $E_m$ with a precision of $10^{-5}$~eV. Detailed calculations show that the EB-HCI scenario becomes realistic using an electron beam ion trap (EBIT).

A simplified EB nuclear excitation scheme is presented in Fig.~\ref{EB_scheme}. The electronic shell, previously brought into an excited state, is further driven by an optical or UV laser. Via a virtual electronic state, EB transfers the excitation energy of the electronic shell to the nucleus. EB is allowed for a particular value of the laser photon energy $\hbar\omega$ defined by the actual isomer energy $E_m$ and the energies of the initial and final electronic levels, $E_i$ and $E_f$, respectively. When a  real electronic state happens to be close to the virtual state, the EB rate can become very large. Thus,  calculated as a function of the isomer energy $E_m$, 
the excitation rate displays a sharp enhancement at the energy which allows the virtual state to be close to a real electronic state.  Once the nuclear excitation process is observed at a particular laser frequency $\omega$, the energy of the isomer can be obtained as $E_m=\hbar\omega+E_i-E_f$.

\begin{figure}[ht!]
\centering
\includegraphics[width=0.47\textwidth]{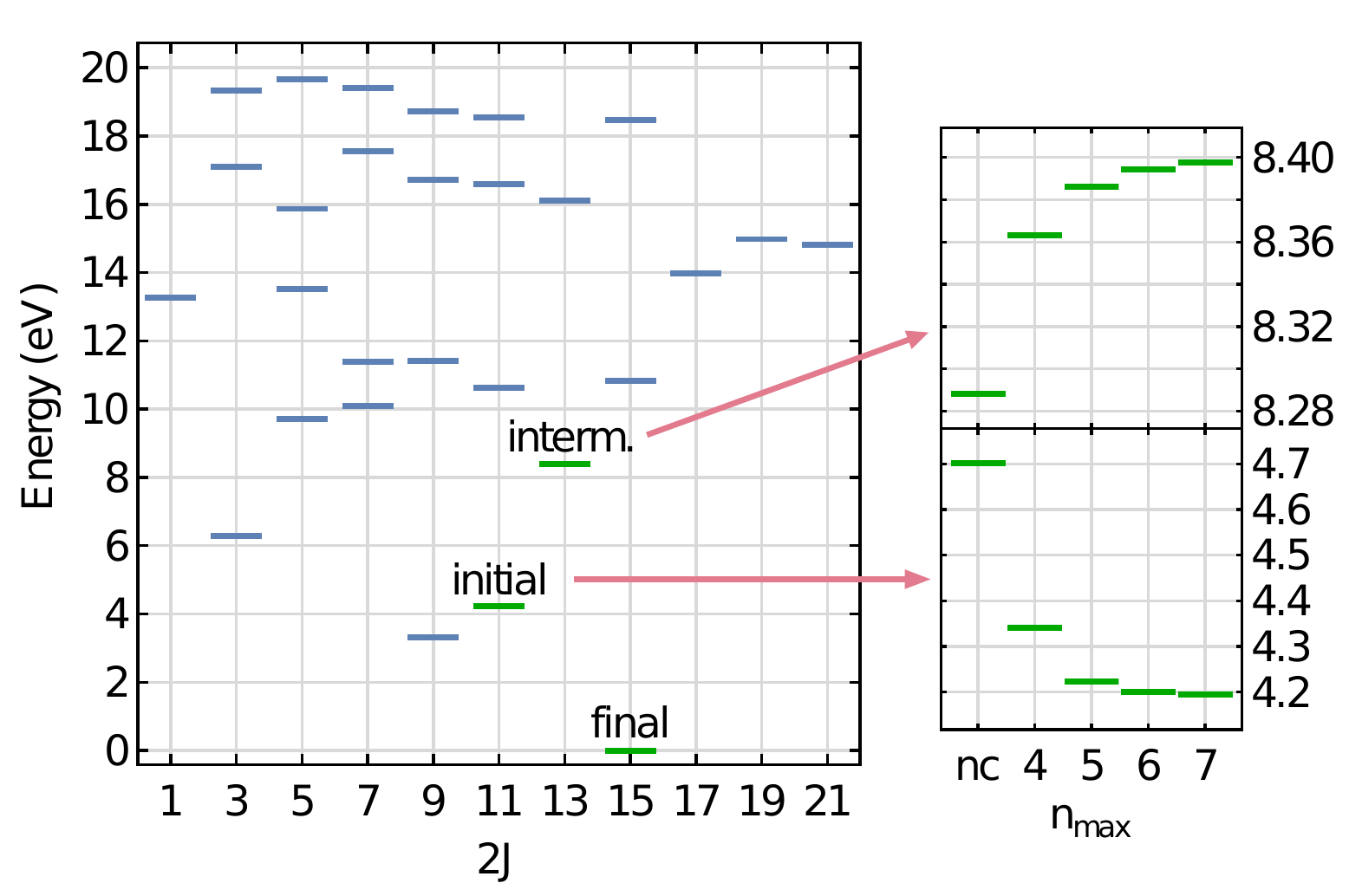}
 \caption{Calculated low-lying electronic energy levels of the $\isotope{Th}^{35+}$ ion and their angular momentum quantum number $J$. The EB states are explicitly labeled. Inset on the right: energies of the two excited states $E_i$ and $E_k$ and their dependence on core-valence electron correlations with excitations up to principal quantum number $n_\text{max}$. The case without correlations is labeled ``nc". See text for explanations. }
\label{spectrum}
\end{figure}

Since no experimental information on the electronic spectra of $\isotope{Th}^{35+}$ is available at present, we rely on \textit{ab initio} calculations with the package GRASP2K \cite{grasp} based on a fully relativistic multiconfiguration Dirac-Hartree-Fock approach. The resulting low-lying electronic levels of interest are shown in Fig.~\ref{spectrum}, with the initial and final electronic states involved in the EB process labeled correspondingly to their role. With {\it intermediate} state we denote the real state closest to the virtual level depicted by the dashed line in Fig~\ref{EB_scheme}. Atomic level energies are sensitive to the inclusion of core-valence electron correlations in the calculation. We take this into account by including double excitations of the active electrons with principal quantum number $n=4$ to orbitals with $n$ up to $n_\text{max}$. For the three relevant energy levels, we choose $n_\text{max}=7$. We have ensured that this choice is sufficient for the purposes of this work by studying the behavior of these energies with $n_\text{max}$ (see inset in Fig.~\ref{spectrum}). The other levels shown are obtained including up to $n_\text{max}=5$ correlations.
In the present work we use the following calculated values: $E_i=4.19$~eV for the initial state  with total angular momentum $J=11/2$, and $E_k=8.40$~eV for the intermediate state with $J=13/2$.

For evaluating the EB excitation rate $\Gamma_\text{excit}$ we start from an idealized case, with the laser tuned to the EB resonance and no Doppler broadening, Zeeman or hyperfine magnetic splitting.
Since the isomer energy $E_m$ is known today only within a $1\sigma$ interval of 0.34~eV, we calculate the dependence of the EB nuclear excitation rate for $E_m$ in that range. We follow Ref. \cite{PorsevFlambaum_Brige_PRL_2010} and express $\Gamma_\text{excit}$ through the rate of the inverse spontaneous process $\Gamma_\text{spont}$ as in Ref. \cite{Sobelman_book_1979}
\begin{equation}\label{EB_EBRatesExcitM}
\Gamma_\text{excit}=\Gamma_\text{spont} \frac{\pi^2 c^2}{\hbar \omega^3} P_\omega \delta\;,
\end{equation}
where $c$ is the speed of light, $P_\omega$ is the laser spectral intensity and $\delta$ depends on the nuclear spin $I_g$ ($I_m$) in the ground (isomeric) state and the electronic shell angular momentum $J_i$ ($J_f$) in the initial (final) state of the EB process as $\delta = (2 I_m+1)(2J_f+1)/[(2 I_g+1)(2J_i+1)]$. Please note that the laser spectral intensity $P_\omega$ is defined here as laser intensity $dI$ per angular frequency interval $d\omega$, whereas the definition in Ref. \cite{Sobelman_book_1979} involves an additional factor $1/(4\pi)$. The rate for the spontaneous inverse EB process can be evaluated following Ref.~\cite{PorsevFlambaum_Brige3+_PRA_2010} with the difference that the absorption of the laser photon is of type $M1$ instead of $E1$ and thus is described by the magnetic moment operator $\hat{\textbf{M}}=\mu_B(\hat{\textbf{L}}+2\hat{\textbf{S}})$, where $\hat{\textbf{L}}$ and $\hat{\textbf{S}}$ are the orbital momentum and spin operators, respectively, and $\mu_B$ is the Bohr magneton. We obtain (in atomic units)
\begin{equation}\label{EB_EBRatesM}
\Gamma_\text{spont}=\frac{4 \pi \omega^3}{3 c^5 (2J_f+1)(2J_k+1)} \sum_{\lambda L} \frac{B_\downarrow^{\lambda L} G^{\lambda L}}{(2L+1)^2}\;,
\end{equation}
where $B_\downarrow^{\lambda L}$ is the reduced probability of the nuclear transition \cite{Ring1980} of the type $\lambda L$ (in our case $M1$ or $E2$) and $G^{\lambda L}$ is given by
\begin{equation}\label{EB_g1M}
G^{\lambda L} = \left| 
 \frac{\rbra{\beta_i J_i}\hat{L}+2\hat{S}\rket{\beta_k J_k} \rbra{\beta_k J_k} \cf{T}_{\ttau L} \rket{\beta_f J_f}}
 {E_k-E_f-E_m}
 \right|^2\, ,
\end{equation}
where $J_k$ and $E_k$ are the total angular momentum and the energy of the intermediate real electronic level, respectively, $\cf{T}_{\ttau L}$ is the electronic coupling operator and the generic indices $\beta$ stand for the quantum numbers which together with the angular momentum describe the electronic states. In third order perturbation theory, the presence of the virtual state requires a summation over all possible electronic states $k$ appearing in the denominator of Eq.~(\ref{EB_g1M}). Here, the largest contribution stems from the real electronic level at 8.40~eV closest to the virtual state and we thus restrict our investigation to this term.

The electronic reduced matrix elements and energies in Eq.~(\ref{EB_g1M}) are obtained numerically using the  GRASP2K package \cite{grasp}. 
For the reduced nuclear transition probabilities entering Eq.~(\ref{EB_EBRatesM}) we assume the values $B_\downarrow^{M1}=0.005$~W.u. and $B_\downarrow^{E2}=29$~W.u. according to the calculations in Ref. \cite{Minkov_Palffy_PRL_2019}, where W.u. denote Weisskopf units~\cite{Ring1980}. The initial state with $E_i=4.19$~eV can be efficiently populated by electron collisions in the EBIT as discussed below. For the tunable UV laser we assume a 30~mJ pulse energy at 100~$\text{s}^{-1}$ pulse-repetition rate and a $2~\text{ GHz}$  linewidth (corresponding to the angular frequency of $2\pi\times 2~\text{ GHz}$) achievable by a standard commercially available short-pulse dye laser pumped at 532~nm and subsequently frequently-doubled to generate 320~nm (approximately 4~eV) wavelength. We find for the EB excitation rate per ion as a function of the isomer energy a resonance at $E_m=8.40$~eV (corresponding to $\hbar\omega=4.21$~eV). While at the low-energy edge of the studied interval $E_m=8.11$~eV the rate yields $9\cdot 10^{-5}$~s$^{-1}$, closer to the resonance this value raises to $7\cdot 10^{-2}$~s$^{-1}$ for $E_m=8.39$~eV. We note that the EB excitation rates per ion are over the entire studied range orders of magnitude larger than the predicted direct photoexcitation rate per ion of $\approx 5\cdot 10^{-10}$~s$^{-1}$ using a VUV laser \cite{Wense_PRL_2017}. 

Next, we study the EB excitation rate for a cloud of trapped $\isotope{Th}^{35+}$ ions under typical EBIT conditions. Loading of rare isotopes in an EBIT has been shown to be feasible by sputtering or laser ablating from a wire probe \cite{Schweiger_RSI_2019, EBIT_ELLIOTT_1995}. The electron beam at the energy $E_0\sim 1$~keV produces $\isotope{Th}^{35+}$ ions and traps them in the radial direction through the Coulomb attraction, whereas three cylindrical electrodes serve for the longitudinal confinement. A magnetic field of $B_0=8$~T generated by superconductive coils compresses the electron beam radially. An advantage of this approach is that while many electronic states with energies up to $E_0$ are constantly populated by the incident electron beam and quickly decay, nearly the entire steady-state population is shared between the ground  and few low-lying metastable states. The latter  are candidates for the initial electronic state in the EB scheme. Simulations performed with the FAC package \cite{FAC} show that the initial level at $E_i=4.19$~eV  chosen in this work has a relative steady-state population of $p_i=17\%$.

\begin{figure}[ht!]
\centering
\includegraphics[width=0.5\textwidth]{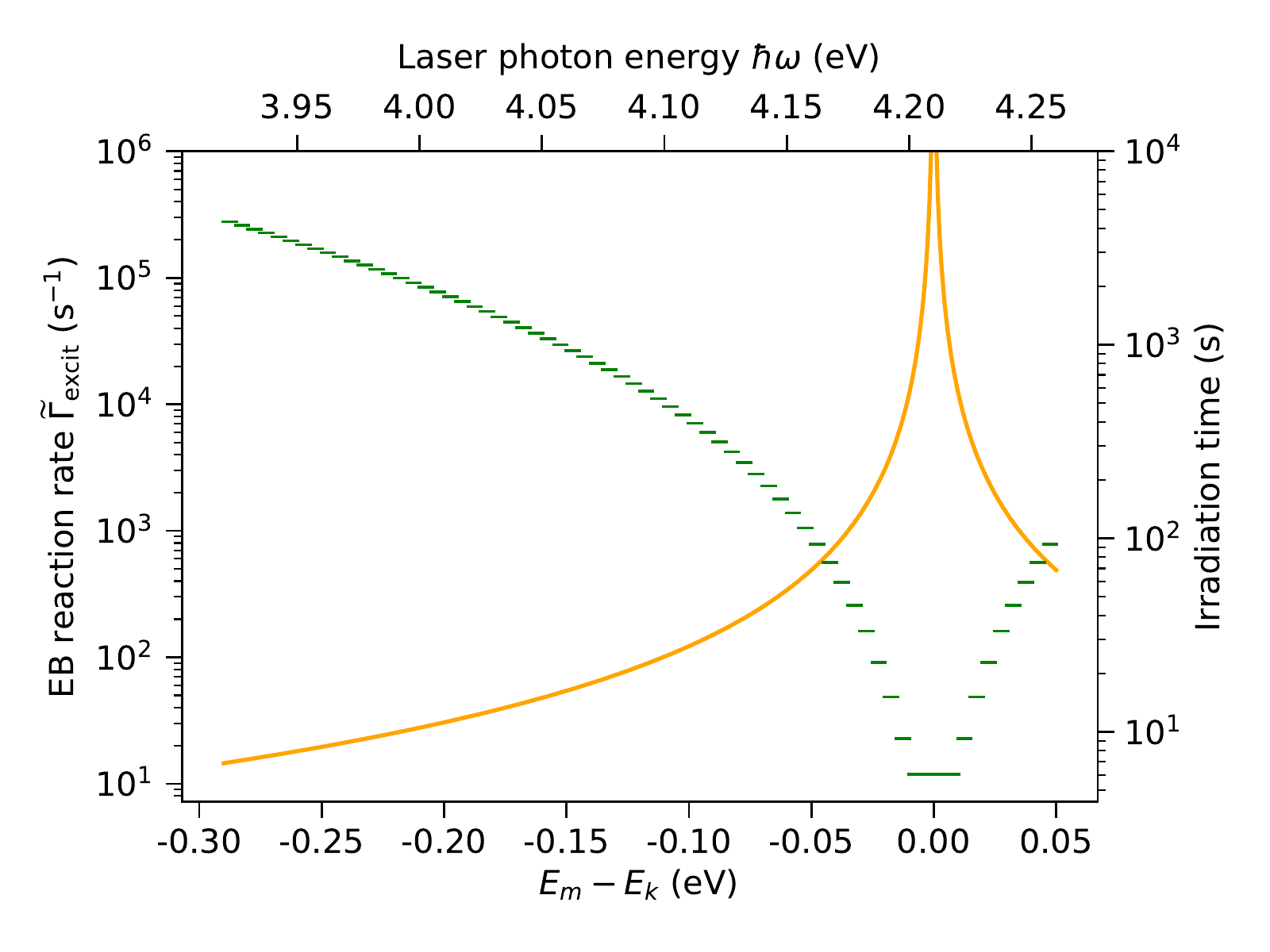}
 \caption{EB reaction rate $\tilde{\Gamma}_\text{excit}=p_i N_\text{HCI} \Gamma_\text{excit}$ (left axis, yellow line) and required exposure time of each 0.005~eV (1.2~THz) bin (right axis, green horizontal segments) for  $\isotope{Th}^{35+}$ ions as a function of the unknown energy difference  $E_m-E_k$  (lower axis), and corresponding laser photon energy (upper axis). }
\label{EB35+}
\end{figure}

Inside the trapping region, we consider an ion cloud with diameter $100\text{ }\mu\text{m}$ containing $N_\text{HCI}=10^6$ $\isotope{Th}^{35+}$ ions irradiated by the UV laser with the time-averaged spectral power $P_\omega=0.03\;\frac{\text{W}\cdot\text{s}}{\text{m}^2}$ corresponding to the introduced laser parameters (30 mJ pulse energy, rectangular pulse shape, 100~$\text{s}^{-1}$ repetition rate, 2~GHz linewidth). The laser scans an energy range of 0.34~eV  corresponding to the presently known $1\sigma$ interval for $E_m$ \cite{Seiferle_Nature_2019}.  At a particular photon energy $\hbar \omega$, the EB process is induced and the nuclear isomeric state is populated. By transferring the HCI after a certain exposure time from the EBIT to an isomer decay detection setup, those ions that have successfully undergone EB can be detected. Subsequent narrowing of the scanning range $\hbar \Delta  \omega$ should allow us to determine the resonant laser frequency and thus the isomer energy $E_m$.
 Figure~\ref{EB35+} shows the calculated EB reaction rate $\tilde{\Gamma}_\text{excit}=p_i N_\text{HCI} \Gamma_\text{excit}$ (prior to the extraction) as a function of the nuclear isomer energy $E_m$ in the range $8.28 \pm 0.17$~eV from Ref.~\cite{Seiferle_Nature_2019}. We use the energy difference $E_m-E_k$  on the lower horizontal axis, and the   absorbed laser photon energy $\hbar\omega$  on the upper horizontal axis. Also the EB reaction rate presents a resonance at $E_m= 8.40$~eV corresponding to the virtual state approaching the chosen intermediate state with energy $E_k$. At resonance, the signal reaches more than $10^5$ counts per second, while the lowest value of 10~s$^{-1}$ would be still detectable even if the nuclear transition lies at 8.11~eV corresponding to $E_m-E_k=-0.29$~eV.

A spectroscopic detection of the isomer based on the hyperfine structure has been recently demonstrated \cite{ThielkingNature_2018}. This approach however cannot be used in the present work due to the Doppler broadening in the EBIT. Instead we employ the detection scheme based on the IC electrons applied in Refs.~\cite{Seiferle_Nature_2019, Wense_Nature_2016}. After a suitable exposure time to the laser, the HCI are extracted from the trap and neutralized. This opens the IC decay channel with 7~$\mu\text{s}$ lifetime \cite{Seiferle_PRL_2017}. Two possible neutralization and detection scenarios are envisaged:
({\bf i}) The ions can be collected on a metallic surface where they neutralize \cite{Wense_Nature_2016, Seiferle_PRL_2017}. The mechanism of HCI neutralization on metallic surfaces is described in Ref.~\cite{HCI_ARNAU1997113}.  The subsequently emitted IC electrons  can be accelerated onto an MCP detector to reach optimum detection efficiency. We estimate a geometrical collection efficiency of 50\%. 
Secondary electrons generated from ionic impact can be separated from IC electrons due to the relatively large IC lifetime.
({\bf ii}) The ions can be neutralized by passing through an ultra-thin carbon foil \cite{Seiferle_Nature_2019}. Neutralization of HCI in graphene was studied in Ref.~\cite{Gruber_NatCommun_2016} in particular for $\isotope{Xe}^{35+}$ for which the characteristic time  $\tau_\text{n}^\text{exp}=2.1\text{ fs}$ for the exponential charge transfer  was measured. For a single graphene layer, this value converts to the incident ion velocity of approx. $1.7$~$\mbox{\normalfont\AA}$/fs corresponding to a kinetic energy of approximately 35~keV for $\isotope[229]{Th}^{35+}$ ions. For efficient neutralization, a low kinetic energy is an easy-to-fulfill condition. Based on our previous experiment \cite{Seiferle_Nature_2019}, we anticipate a 50\% neutralization efficiency.  The IC electrons emitted in-flight by the neutral atoms  can be collected in a homogeneous magnetic field along the beam axis and guided to an MCP detector background-free and with close-to-unity efficiency. For both approaches, we further consider a  50~\% MCP detection efficiency \cite{Galanti_1971}, leading to a total detection efficiency of 25\% in both cases.

\begin{figure}[ht!]
\centering
\includegraphics[width=0.45\textwidth]{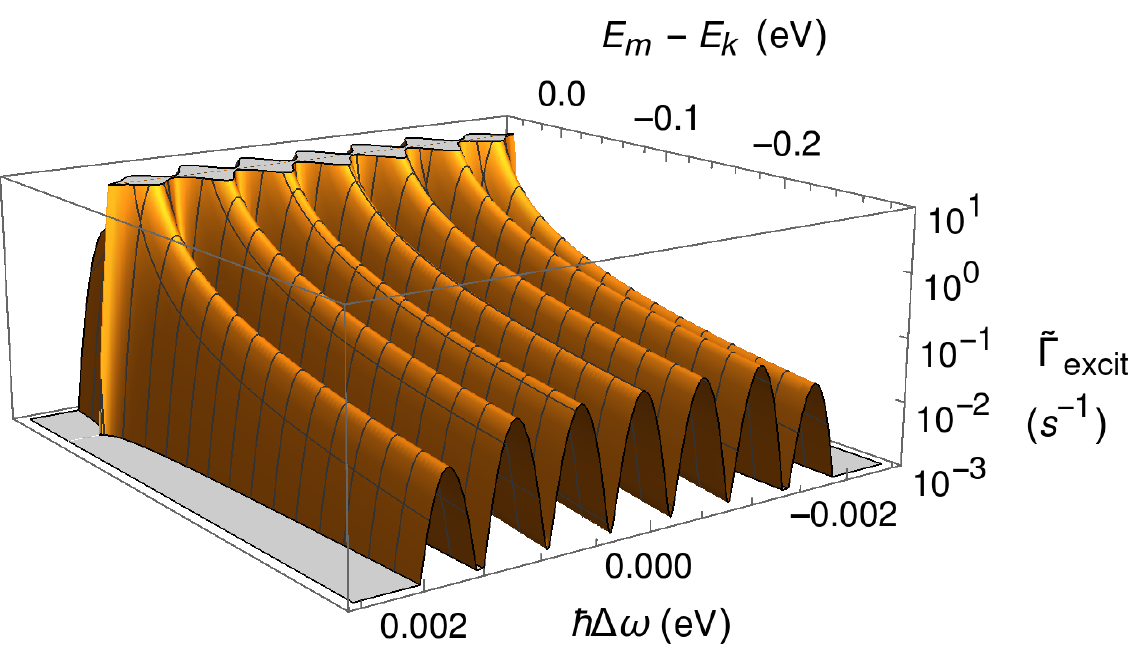}
 \caption{Zeeman splitting and Doppler broadening of the absorbed photon energy. The EB reaction rate is presented as a function of the laser detuning $\hbar \Delta \omega$ from the resonant value and the energy difference $E_m-E_k$,  for  $E_m$ in the recently measured uncertainty interval $8.28 \pm 0.17$~eV \cite{Seiferle_Nature_2019}.}
\label{zeeman}
\end{figure}

In a realistic setup, the accuracy of the isomer energy  determination is limited by several effects: Doppler broadening, Zeeman splitting of electronic levels in the strong EBIT magnetic field, and hyperfine splitting. In the EBIT, the HCI temperature can range from 100 eV down to 10 eV as shown from laser spectroscopy \cite{Maeckel_PRL_2011, Maeckel_PRL_2011_Note, Schnorr_ApJ_2013} and optical line emission measurements \cite{Bekker_PRA_2018}. Figure ~\ref{zeeman} illustrates the calculated rate $\tilde{\Gamma}_\text{excit}$ as a function of both the laser detuning from the resonant value $\hbar \omega$ in Fig.~\ref{EB35+}  and the difference $E_m - E_k$ considering the combined effects of Doppler broadening for 100 eV ion temperature and Zeeman splitting in the EBIT magnetic field of $B_0=8$~T. The Zeeman effect splits the resonance in seven distinctive peaks. The $g$-factors of the involved states are calculated with the GRASP2K package \cite{grasp}. 
The hyperfine splitting is for these parameters smaller than the Doppler width, and thus not resolved. For these experimental parameters, we estimate a $10^{-5}$ eV accuracy for the isomer energy determination.

The isomer energy search within the interval $8.28 \pm 0.17$~eV can be initially performed in 68 bins of 0.005~eV (1.2~THz) width for each excitation and detection loop. At first scanning within one 0.005~eV bin could be performed in approx. 600 subintervals with the size of the laser linewidth (2~GHz). The pumping time in each 2~GHz subinterval should be chosen such that, provided the laser would hit the resonance, the respective calculated EB excitation rate in Fig.~\ref{EB35+} guarantees a clear detection signal. In the following we consider 10 measured isomer signals (taking into account the setup detection efficiency discussed above) as  successful detection requirement. Once the entire 1.2 THz bin has been scanned, the HCI are extracted and enter the detection setup. 
 The procedure is repeated for each of the 68 bins until the isomer excitation  signal is detected. Subsequent narrowing of the scanning range should allow us to precisely determine the  laser frequency corresponding to the EB resonance.

We can now estimate the necessary irradiation time within one 0.005 eV bin as a function of the (unknown) energy difference between the intermediate electronic state and the nuclear isomer. The time intervals are presented on the right-hand axis of Fig.~\ref{EB35+}.  If the isomer energy is close to 8.11~eV, the laser scanning time per bin approaches $5000$~s, which corresponds to approximately 900 shots (in 9~s) in each 2~GHz subinterval. Since here the scanning time is of the same order of magnitude as the predicted isomer lifetime of $ 1.1 \cdot 10^4 $~s in ions, we correct for the radiative decay of the isomers prior to extraction (the spontaneous EB decay channel does not affect the isomer lifetime). Close to the resonance at $E_m=E_k=8.40$~eV, the EB rate is so strong that one single laser shot would cause the observation of 10 isomer excitations.
If all intervals have to be scanned, we estimate a total irradiation time of 22 hours. The final scans for the value of $E_m$ within the obtained 0.005~eV interval can be done much faster and are therefore negligible in the overall duration. The presence of Zeeman splitting may distribute the signal between bins, in which case a finer division should be used. We note that the assumed parameters for the ionic cloud and the laser setup are quite conservative and the scheme can be implemented using an EBIT and a commercially available tunable UV laser. Spectroscopic measurements of the $M1$ emission lines in the optical and VUV range and application of the Ritz-Rydberg method, as in Refs.~\cite{Bekker_NatComm_2019,Windberger_PRL_2015} would be needed for the final determination of the nuclear isomer energy as $E_m=\hbar\omega+E_i-E_f$. Based on previous EBIT experiments, we can expect a relative accuracy of the measured electronic energy levels of $10^{-6}$.

In conclusion, our calculations show that the use of HCI for EB excitation of the Th isomer is very promising for an efficient isomer population as well as for a more precise determination of the isomer energy. The proposed setup comprises an EBIT, a tunable UV laser,  and an electron spectrometer for the detection of the isomer decay through IC in neutralized ions, and is feasible already today. Using HCI, we can profit from an electronic level scheme rich with $M1$ and $E2$ transitions at low energies and from a high laser intensity without multiphoton ionization. Eventually, a successful EB scheme in $\isotope{Th}^{35+}$ should provide more efficient isomer excitation than direct VUV laser photoexcitation and could also potentially be used for the operation of a future nuclear clock. 

PVB, BS, LvdW, PGT and AP gratefully acknowledge funding from the European Union's Horizon 2020 research and innovation
program under grant agreement No 664732 ``nuClock''. The work of HB and JRCLU is part of and supported by the DFG Collaborative Research Center SFB 1225 (ISOQUANT). JCB was supported by the Alexander von Humboldt Foundation.

\section*{References}
\bibliography{refs}

\end{document}